\documentclass[conference]{IEEEtran}
\usepackage[hidelinks]{hyperref}
\IEEEoverridecommandlockouts
\usepackage{cite}
\usepackage{amsmath,amssymb,amsfonts}
\usepackage{tabularx}
\usepackage{makecell}
\usepackage{algorithmic}
\usepackage{graphicx}
\usepackage{arydshln}

\usepackage{textcomp}
\usepackage{xcolor}
\usepackage{booktabs}
\usepackage{multirow}
\usepackage{rotating}

\def\BibTeX{{\rm B\kern-.05em{\sc i\kern-.025em b}\kern-.08em
    T\kern-.1667em\lower.7ex\hbox{E}\kern-.125emX}}

\begin{document}

\title{RFWM: Physics-Guided World Model for Dynamic Wireless Radiance Field Generation}

\author{\IEEEauthorblockN{Zijiu Yang and Qianqian Yang }
\IEEEauthorblockA{ 
College of Information Science and Electronic Engineering, Zhejiang University, Hangzhou, China \\
Email: \{zijiu\_yang, qianqianyang20\}@zju.edu.cn
}

}

\maketitle
\begin{abstract}
Radio-frequency (RF) radiance-field modeling is essential for wireless network optimization and sensing, yet remains challenging in dynamic and unseen environments. Existing learning-based methods synthesize RF fields from sparse measurements, but most struggle to generalize to dynamic and unseen environments. 
To address this limitation, we propose RFWM, a physics-guided RF world model that maps multimodal physical conditions like visual dynamics and AP configurations to spatiotemporal RF fields.
RFWM adopts a two-stage training strategy with physics-guided priors and constraints.
In the first stage, RFWM adapts a pretrained visual diffusion backbone to RF trajectories to predict RF sequences from a few past RF inputs, while conditioning the backbone on a Friis-guided prior for coarse attenuation guidance.
In the second stage, RFWM learns the physical-to-RF mapping by training a ControlNet from scratch and fine-tuning the RF-adapted backbone, while six physics-guided regularizers enforce fine-grained propagation consistency.
Cross-height heads then jointly generate RF trajectories at queried receiver heights in one forward pass.
We construct a new benchmark of 7,715 sequences averaging 33 frames across 115 environments for dynamic RF-field generation. Experimental results show that RFWM improves MSE by approximately 7 dB and 3 dB over the state of the art under in-distribution and out-of-distribution settings, respectively.
\end{abstract}


\section{Introduction}
Wireless connectivity has become essential infrastructure supporting a broad range of modern digital systems and applications, such as mobile computing, intelligent transportation and Internet-of-Things applications\cite{RF-Robot,zeng2024ckm,bi2019radiomaps}. The reliable operation of these systems depends on accurate radio-frequency (RF) field modeling, which characterizes the spatial distribution of received signal strength (RSS) and supports tasks such as network planning, resource allocation, and communication-aware control. Although RF propagation is fundamentally governed by Maxwell's equations \cite{yun2015raytracing},  accurately modeling it in realistic environments requires detailed knowledge of scene geometry, material properties, and boundary conditions. These challenges have motivated three major classes of RF modeling approaches: stochastic channel models, geometry-based propagation models, and learning-based methods.

Stochastic channel models, such as COST 2100 \cite{cost2100} and QuaDRiGa \cite{quadriga}, characterize RF propagation using statistical distributions derived from measurements or simulations of representative environments. Despite their computational efficiency, these models abstract away detailed environmental characteristics, such as scene geometry, material properties, and object configurations, and therefore cannot capture fine-grained, site-specific RF variations. In contrast, geometry-based ray-tracing frameworks such as Sionna RT \cite{sionna} explicitly model propagation paths based on the physical environment, thereby providing greater site-specific fidelity. However, tracing numerous propagation paths incurs substantial computational overhead, requires accurate descriptions of scene geometry and material properties, and must be repeated whenever the environment or wireless configuration changes \cite{yun2015raytracing}.

Learning-based approaches offer a promising alternative by learning propagation patterns directly from data.
Some existing methods, however, adopt a \emph{scene-specific fitting} paradigm, in which model parameters or scene representations are optimized using RF measurements collected in a target environment.
For example, NeRF$^2$ \cite{zhao2023nerf2} and NeWRF \cite{lu2024newrf} use implicit neural fields to reconstruct wireless propagation from sparse RF measurements, while WRF-GS \cite{wen2025wrfgs} and RadCloudSplat \cite{wang2026radcloudsplat} employ explicit Gaussian representations to accelerate RF-field rendering and radiomap extrapolation. 
The scene-specific nature of these methods prevents their learned propagation representations from generalizing directly to previously unseen environments. Furthermore, their reconstructed RF fields become outdated when environmental changes occur, such as the movement of people or objects \cite{RadioUNet,SURM}.

More recent methods have begun to address environmental dynamics and cross-scene generalization. 
RFCanvas \cite{chen2024rfcanvas} updates generated RF fields using motion cues extracted from visual observations, whereas 
GRaF \cite{yang2026generalizable} predicts the spatial spectrum for a new AP placement using RF measurements collected at nearby AP locations in the same scene.
Both methods move beyond purely static reconstruction by improving adaptation to environmental changes or spatial generalization.
However, these methods still represent environmental changes as discrete configurations rather than continuous temporal evolution, and remain dependent on target-scene RF measurements.
Diffusion$^2$ \cite{park2026diffusion} further advances dynamic RF modeling by conditioning on frame-wise 3D scene snapshots to generate RF heatmap videos across scenes. Nevertheless, it requires frame-wise 3D geometry and sparse RF measurements 
as input and is limited to generating single-height 2D RF heatmaps.

These limitations motivate a fundamental question: \emph{can a shared model transfer propagation knowledge to an unseen environment and predict the temporal evolution of its 3D RF field without any target-scene RF measurements?} Recent advances in \textit{visual world models} suggest a promising path toward this goal. By learning spatial structures and temporal dynamics from sequential observations, world models can predict future states and transfer learned dynamics across environments \cite{hafner2023dreamerv3,bruce2024genie}. One representative example is Cosmos, a family of world foundation models pretrained on large-scale physical-world videos to acquire transferable spatial and temporal priors \cite{agarwal2025cosmos}. 
Building on these priors, Cosmos-Transfer \cite{alhaija2025cosmostransfer} introduces \emph{world-to-world transfer}, in which structural and temporal cues from one modality, such as depth or segmentation, guide the generation of another modality. This paradigm provides a shared-model foundation for transferring physical structure and dynamics across representations.

Inspired by this, we formulate dynamic RF-field generation as a \emph{physical-to-RF world transfer} problem, where a shared model maps physical-world observations, including scene dynamics, static geometry, and wireless deployment, to spatiotemporal RF fields. To realize this formulation, we propose \textbf{RFWM}, a physics-guided RF world model that enables direct RF-field generation in both seen and unseen environments. RFWM employs a two-stage training framework. In the first stage, it adapts a pretrained visual world-model backbone \cite{agarwal2025cosmos,dit} to predict future RF states from historical RF trajectories and a propagation prior condition while preserving the spatiotemporal knowledge acquired from physical-world videos. The propagation prior is derived from the Friis free-space path-loss equation \cite{frisis} and provides coarse guidance on large-scale signal attenuation.
In the second stage, the RF-adapted backbone learns physical-to-RF generation by incorporating multimodal conditions \cite{controlnet}, such as visual dynamics and access-point (AP) deployment, through dedicated pathways designed according to their structures and physical roles. Six physics-guided regularization terms further enforce fine-grained consistency with blockage effects, distance-dependent attenuation, and local spatial structure. RFWM further introduces cross-height heads to efficiently model the vertical structure of 3D RF fields. These heads share backbone features across receiver heights while preserving height-specific propagation characteristics. Unlike existing methods that predict a single RSS value per inference \cite{zhao2023nerf2}, RFWM jointly generates complete spatiotemporal RF fields at all queried receiver heights in a single forward pass. 

Our main contributions are summarized as follows:
\begin{itemize}
\item We formulate dynamic RF-field modeling as a \emph{physical-to-RF world transfer} problem, in which a shared model maps physical-world observations to spatiotemporal RF fields and directly generalizes to unseen environments without target-scene RF measurements.

\item We propose \textbf{RFWM}, a physics-guided RF world model with a two-stage training framework. Stage~I adapts a pretrained visual world-model backbone to predict future RF states from historical RF trajectories and a Friis-guided propagation prior. Stage~II conditions the RF-adapted backbone on multimodal physical observations for physical-to-RF generation, while six physics-guided regularizers enforce consistency with blockage effects, distance-dependent attenuation, and local spatial structure.

\item We develop an efficient multi-height 3D RF generation scheme based on cross-height heads. By sharing backbone features while retaining height-specific propagation characteristics, RFWM generates complete spatiotemporal RF fields at all queried heights in a single forward pass.

\item We construct a multi-scene benchmark containing 7,715 dynamic RF sequences from 115 environments. Extensive experiments show that RFWM outperforms the state of the art by approximately \(7\) dB on in-distribution (ID) scenes and \(3\) dB on out-of-distribution (OOD) scenes, demonstrating strong generalization to unseen environments.

\end{itemize}

\section{Preliminaries}
\label{sec:preliminaries}
This section introduces the RF-field representation used throughout the paper, and formulates dynamic RF generation as a physical-to-RF world-transfer problem.

\subsection{Dynamic Spatiotemporal RF Fields}
\label{subsec:dynamic_rf_field}

Let $\mathcal{S}$ denote a collection of wireless environments, with $s\in\mathcal{S}$ indexing one environment. Its three-dimensional spatial domain is $\mathcal{V}_s\subseteq\mathbb{R}^{3}$, and a trajectory contains $T$ time steps indexed by $t\in\{1,\ldots,T\}$. We write the physical state at time $t$ as
\begin{equation}
\mathbf{x}_{s,t}
=
\bigl(\mathbf{g}_s,\mathbf{d}_{s,t}\bigr),
\label{eq:physical_state}
\end{equation}
where $\mathbf{g}_s$ describes the time-invariant scene structure and $\mathbf{d}_{s,t}$ describes time-varying humans and movable objects. The corresponding physical trajectory is $\mathbf{X}_s=(\mathbf{x}_{s,t})_{t=1}^{T}$.

The transmitter configuration is denoted by
\begin{equation}
\boldsymbol{\eta}_s
=
\bigl(
\mathbf{p}^{\mathrm{tx}}_s,
\mathbf{o}^{\mathrm{tx}}_s,
f_{c,s},
P^{\mathrm{tx}}_s
\bigr),
\label{eq:tx_configuration}
\end{equation}
where $\mathbf{p}^{\mathrm{tx}}_s$ and $\mathbf{o}^{\mathrm{tx}}_s$ are the transmitter position and orientation, $f_{c,s}$ is the carrier frequency, and $P^{\mathrm{tx}}_s$ is the transmit power. For a receiver at $\mathbf{r}\in\mathcal{V}_s$, let $h_{s,t}(\mathbf{r})$ denote the aggregate complex channel induced by the physical state and transmitter configuration. The received signal strength (RSS) is
\begin{equation}
\rho_{s,t}(\mathbf{r})
=
10\log_{10}\!\left(
\frac{P^{\mathrm{tx}}_s\lvert h_{s,t}(\mathbf{r})\rvert^2}{P_0}
\right),
\label{eq:rss}
\end{equation}
where $P_0$ is a reference power. Motion in $\mathbf{d}_{s,t}$ changes blockage, path visibility and multipath interactions, making the RF fields evolve over time.

For learning and evaluation, we sample the RF field on an $H\times W$ horizontal grid
\begin{equation}
\mathcal{G}_s
=
\left\{
(x_i,y_j)
\mid
i=1,\ldots,H,\;j=1,\ldots,W
\right\}
\end{equation}
and at receiver heights $\mathcal{Z}=\{z_k\}_{k=1}^{K}$. The RF field at time $t$ is
\begin{equation}
\mathbf{R}_{s,t}[k,i,j]
=
\rho_{s,t}(x_i,y_j,z_k),
\qquad
\mathbf{R}_{s,t}\in\mathbb{R}^{K\times H\times W}.
\label{eq:sampled_rf_field}
\end{equation}

To leverage the pretrained visual VAE for compact latent encoding of RF tokens, we convert each RSS field into
an RGB representation using a fixed rendering operator \(\Psi(\cdot)\).
For receiver height \(z_k\), let
\(r_{s,t,k}(i,j)=\mathbf{R}_{s,t}[k,i,j]\) denote the RSS at grid
location \((i,j)\). The rendering process is defined as
\begin{equation}
\begin{aligned}
u_{s,t,k}(i,j)
&=
\operatorname{clip}
\left(
\frac{r_{s,t,k}(i,j)-r_{\min}}
     {r_{\max}-r_{\min}},
0,1
\right), \\
\left[\mathbf{Y}_{s,t}\right]_{k,:,i,j}
&=
\mathbf{J}_{\mathrm{jet}}
\left(
u_{s,t,k}(i,j)
\right), \\
\mathbf{Y}_{s,t}
&=
\Psi\left(\mathbf{R}_{s,t}\right)
\in
[0,1]^{K\times 3\times H\times W},
\end{aligned}
\label{eq:rendered_rf_field}
\end{equation}
where \(u_{s,t,k}(i,j)\in[0,1]\) is the normalized RSS value, and
\(\mathbf{J}_{\mathrm{jet}}:[0,1]\rightarrow[0,1]^3\) maps it to RGB.
We fix \(r_{\min}=-70\,\mathrm{dB}\) and
\(r_{\max}=-30\,\mathrm{dB}\) for all scenes and time steps. 
Since \(\Psi(\cdot)\) is applied pointwise,
it changes only the value representation while preserving the temporal,
spatial, and height structure of the RF field.

The complete dynamic 3D RF-field trajectory is obtained by stacking the
rendered fields over time:
\begin{equation}
\mathbf{Y}_s
=
\left(
\mathbf{Y}_{s,t}
\right)_{t=1}^{T}
\in
[0,1]^{T\times K\times 3\times H\times W}.
\label{eq:dynamic_rf_trajectory}
\end{equation}
For the height-wise latent modeling used later, we further define
\begin{equation}
\begin{aligned}
\mathbf{R}_{s,k}
&=
\left(
\mathbf{R}_{s,t}[k,:,:]
\right)_{t=1}^{T}
\in
\mathbb{R}^{T\times H\times W}, \\
\mathbf{Y}_{s,k}
&=
\Psi\left(
\mathbf{R}_{s,k}
\right)
=
\left(
\mathbf{Y}_{s,t}[k,:,:,:]
\right)_{t=1}^{T}
\in
[0,1]^{T\times 3\times H\times W}.
\end{aligned}
\label{eq:heightwise_rf_trajectory}
\end{equation}
Accordingly, \(\mathbf{Y}_s\) is obtained by stacking
\(\{\mathbf{Y}_{s,k}\}_{k=1}^{K}\) along the receiver-height dimension and
jointly represents temporal evolution, horizontal propagation structure and
cross-height variation.

\subsection{Physical-to-RF World Transfer}
\begin{figure}[t]
    \centering
    \includegraphics[width=\columnwidth]{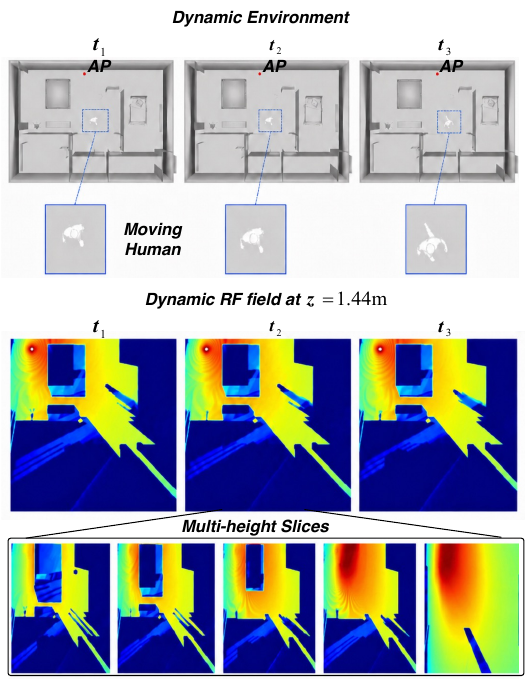}
   \caption{Illustration of synchronized physical and RF observations in a dynamic environment.}
    \label{fig1}
\vspace{-4mm}
\end{figure}
Unlike scene-specific fitting methods
\cite{zhao2023nerf2,lu2024newrf,wen2025wrfgs}, RFWM treats the observable
physical world and the resulting RF field as two synchronized representations
of the same dynamic environment, as shown in \autoref{fig1}.
Let \(\mathbf{Y}_s\) denote the spatiotemporal RF field of scene
\(s\), and let \(\mathbf{C}_s^{\mathrm{obs}}\) denote the physical observations
available for that scene. We formulate physical-to-RF world transfer as
learning
\begin{equation}
    p_{\boldsymbol{\theta}}
    \left(
        \mathbf{Y}_s
        \mid
        \mathbf{C}_s^{\mathrm{obs}}
    \right),
    \label{eq:physical_to_rf_transfer}
\end{equation}
where a single parameter set \(\boldsymbol{\theta}\) is shared across diverse environments. 

Specifically, the physical observations for scene \(s\) are organized as
\begin{equation}
\mathbf{C}_s^{\mathrm{obs}}
=
\left(
\mathbf{V}_s,
\mathbf{M}_s,
\mathbf{e}_s,
\mathbf{A}_s,
\mathbf{W}_s
\right),
\label{eq:rfwm_condition}
\end{equation}
where the synchronized visual sequence \(\mathbf{V}_s\) records the motion of humans
and movable objects, the static geometry \(\mathbf{M}_s\) describes the
3D structure of the environment and the text embedding \(\mathbf{e}_s\) provides global semantic context.
$\mathbf{A}_s$ stands for the AP configurations with the transmitter position and horizontal orientation encoded as a spatial
raster aligned with the visual and RF observations, as expressed by
\begin{equation}
\mathbf{A}_s
=
\mathcal{R}_{\mathrm{AP}}
\left(
\mathbf{p}_s^{\mathrm{tx}},
\mathbf{o}_s^{\mathrm{tx}}
\right)
=
\left[
\mathbf{A}_s^{\mathrm{pos}},
\mathbf{A}_s^{\cos},
\mathbf{A}_s^{\sin}
\right],
\label{eq:ap_raster}
\end{equation}
where \(\mathcal{R}_{\mathrm{AP}}(\cdot)\) denotes the rasterization operator.
\(\mathbf{A}_s^{\mathrm{pos}}\) is a Gaussian map centered at the
transmitter location, while \(\mathbf{A}_s^{\cos}\) and
\(\mathbf{A}_s^{\sin}\) encode the cosine and sine of its horizontal
orientation, respectively. 
$\mathbf{W}_s$ is organized over the \(K\) queried receiver
heights as
\begin{equation}
\begin{aligned}
\mathbf{W}_s
&=
\left(
\mathbf{w}_{s,k}
\right)_{k=1}^{K}, 
\mathbf{w}_{s,k}
=
\left[
B_s,
f_{c,s},
z_k,
P_s^{\mathrm{tx}},
\boldsymbol{\delta}_s,
\boldsymbol{\beta}_s
\right],
\end{aligned}
\label{eq:wireless_condition}
\end{equation}
where \(B_s\), \(f_{c,s}\), and \(P_s^{\mathrm{tx}}\) denote the system
bandwidth, carrier frequency, and transmit power, respectively, and \(z_k\)
denotes the \(k\)-th queried receiver height. Moreover,
\(\boldsymbol{\delta}_s\) specifies the metric scale of the 3D scene, while
\(\boldsymbol{\beta}_s\) contains the material parameters that characterize
reflection behavior.

Conditioned on these observations, RFWM generates the corresponding RF field
$\widehat{\mathbf{Y}}$ according to
\begin{equation}
    \widehat{\mathbf{Y}}_s
    =
    F_{\boldsymbol{\theta}}
    \left(
        \mathbf{C}_s^{\mathrm{obs}}
    \right),
    \label{eq:physical_to_rf_generation}
\end{equation}
where $F_{\boldsymbol{\theta}}(\cdot)$ represents the transfer operator.

\begin{figure*}[t]
    \vspace{-4mm}
    \centering
    \includegraphics[width=\textwidth]{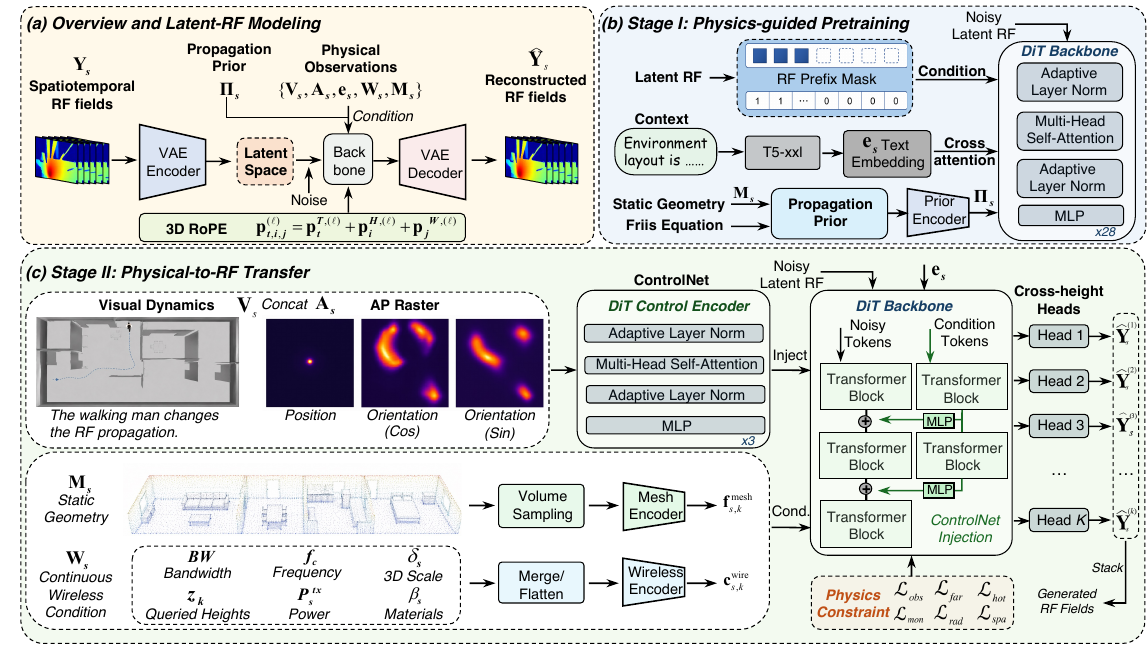}
    \vspace{-5mm}
    \caption{Overview of RFWM. (a) Spatiotemporal RF fields are
    rendered and compressed into latent RF trajectories using a pretrained VAE,
    while a DiT backbone equipped with 3D RoPE models their spatial and temporal
    evolution. (b) Stage~I adapts the pretrained visual DiT to the RF domain
    through RF-prefix conditioning, scene-level text cross-attention, and a
    Friis-guided propagation prior. (c) Stage~II learns the physical-to-RF
    mapping from deployment-time physical observations. Cross-height heads
    jointly generate RF fields at all queried receiver heights, while six
    physics-guided regularization terms improve fine-grained physical
    consistency.}
    \label{fig2}
    \vspace{-4mm}
\end{figure*}

\section{RFWM}
\label{sec:rfwm}
RFWM realizes physical-to-RF world transfer with a latent diffusion backbone
initialized from a pretrained visual diffusion-based world model \cite{alhaija2025cosmostransfer}. 
While visual pretraining
provides general spatiotemporal modeling capabilities, it does not capture the
propagation-specific characteristics of RF fields. RFWM therefore separates RF-domain
adaptation from physical-to-RF conditioning with the 
two-stage training design as shown in \autoref{fig2}. 
Stage~I performs physics-guided pretraining on RF trajectories to learn a shared RF-aware backbone and then Stage~II learns the physical-to-RF mapping.

\subsection{Latent-RF Modeling and Overview}
\label{subsec:rfwm_overview}

\subsubsection{Latent RF Representation}
Directly generating RF trajectories would require the
Diffusion Transformer (DiT) backbone to process a prohibitively dense set of
spatiotemporal RF tokens.
RFWM instead leverages a pretrained visual VAE to tokenize the RF trajectories $\mathbf{R}_{s,t}$ into a compact latent space.

For each height, the rendered trajectory is encoded into the VAE latent space
and perturbed at diffusion noise level \(\sigma\):
\begin{equation}
\begin{aligned}
\mathbf{z}_{0,s,k}
&=
\mathcal{E}_{\mathrm{VAE}}
\left(
\mathbf{Y}_{s,k}
\right)
\in
\mathbb{R}^{
T_{\mathrm{lat}}
\times C_{\mathrm{lat}}
\times H_{\mathrm{lat}}
\times W_{\mathrm{lat}}
},
\\
\mathbf{z}_{\sigma,s,k}
&=
\mathbf{z}_{0,s,k}
+
\sigma\boldsymbol{\epsilon}_{s,k},
\qquad
\boldsymbol{\epsilon}_{s,k}
\sim
\mathcal{N}(\mathbf{0},\mathbf{I}),
\end{aligned}
\label{eq:rf_latent_encoding}
\end{equation}
where $\mathbf{z}_{0,s,k}$ denotes the latent RF representations and $\mathbf{Y}_{s,k}$ is defined in \eqref{eq:heightwise_rf_trajectory}.
All receiver-height planes within the same sample share the same noise level \(\sigma\), while their noise realizations are sampled independently. The resulting 3D latent representation is
$\mathbf{Z}_{\sigma,s}
=
\operatorname{Stack}
\left(
\mathbf{z}_{\sigma,s,1},
\ldots,
\mathbf{z}_{\sigma,s,K}
\right).$

After denoising, the estimated clean latent at each height is decoded to reconstruct the corresponding rendered RF sequence:
$\widehat{\mathbf{Y}}_{s,k}
=
\mathcal{D}_{\mathrm{VAE}}
\left(\widehat{\mathbf{z}}_{0,s,k}\right).$
Finally, the rendered RF sequences from all \(K\) receiver heights are stacked to form the complete spatiotemporal RF field as
$\widehat{\mathbf{Y}}_s
=
\operatorname{Stack}
\left(
\widehat{\mathbf{Y}}_{s,1},
\ldots,
\widehat{\mathbf{Y}}_{s,K}
\right).$

\subsubsection{Spatiotemporal Position Encoding}
Each latent token is indexed by \((k,t,i,j)\), corresponding to the receiver height, time step, and two horizontal coordinates, respectively. To model relative dependencies within each RF sequence, we apply 3D rotary position encoding \cite{3drope} over the temporal and horizontal axes \((t,i,j)\) to the attention queries and keys. In addition, each transformer block incorporates separable learned embeddings for all four axes, as expressed by
\begin{equation}
\mathbf{p}^{(\ell)}_{k,t,i,j}
=
\mathbf{p}^{Z,(\ell)}_{k}
+
\mathbf{p}^{T,(\ell)}_{t}
+
\mathbf{p}^{H,(\ell)}_{i}
+
\mathbf{p}^{W,(\ell)}_{j},
\label{eq:rf_position_embedding}
\end{equation}
where \(\ell\) indexes the DiT block. The rotary encoding captures relative spatiotemporal displacements while the learned embeddings provide absolute height, time, and spatial information.

\subsection{Physics-Guided RF Pretraining}
\label{subsec:rf_pretraining}
As shown in \autoref{fig2}~(b), Stage~I performs physics-guided RF pretraining to adapt the pretrained visual DiT backbone to the RF domain. The backbone is conditioned on three complementary cues: an optional RF prefix, the global text embedding \(\mathbf{e}_s\) encoded by T5-xxl \cite{T5}, and the Friis-guided propagation prior \(\boldsymbol{\Pi}_s\). The RF prefix provides past RF observations for temporal completion, the text embedding injects scene-level semantics through cross-attention, and the propagation prior combines Friis-based attenuation with geometry-aware visibility information.

\subsubsection{RF-Prefix Conditioning}
The RF prefix exposes the backbone to varying amounts of historical RF context.
For each training sample, we draw
$n_{\mathrm{pre}}
\sim
\operatorname{Unif}
\left(
\left\{
0,\ldots,
\left\lfloor
T_{\mathrm{lat}}/3
\right\rfloor
\right\}
\right)$
and define a binary mask
\(\mathbf{B}_{s,k}^{\mathrm{pre}}\) that equals one on the first
\(n_{\mathrm{pre}}\) latent slices and zero elsewhere, broadcast over the
channel and spatial dimensions. A small perturbation is also added to the observed
slices, and the masked values are concatenated with the mask:
\begin{equation}
\mathbf{c}_{s,k}^{\mathrm{pre}}
=
\operatorname{Concat}_{\mathrm{ch}}
\left[
\mathbf{B}_{s,k}^{\mathrm{pre}}
\odot
\left(
\mathbf{z}_{0,s,k}
+
\sigma_c
\boldsymbol{\epsilon}_{c,s,k}
\right),
\mathbf{B}_{s,k}^{\mathrm{pre}}
\right],
\label{eq:rf_prefix_condition}
\end{equation}
where
\(\boldsymbol{\epsilon}_{c,s,k}
\sim\mathcal{N}(\mathbf{0},\mathbf{I})\)
is independent of the diffusion noise and \(\sigma_c\) denotes the conditioning-noise scale.
Importantly, when
\(n_{\mathrm{pre}}=0\),
the backbone is trained to generate RF trajectories without target-scene RF observations,
which facilitates the operating setting of Stage~II.

\subsubsection{Friis-Guided Propagation Prior}
The RF prefix provides data-driven temporal guidance but does not explicitly contain the physical laws governing RF propagation. To complement this, we construct a Friis-guided propagation prior computed with the AP configuration and static scene geometry.
For a receiver at
\(\mathbf{r}_{k,i,j}=(x_i,y_j,z_k)\), the AP--receiver distance is
\begin{equation}
d_{s,k}[i,j]
=
\left\|
\mathbf{r}_{k,i,j}
-
\mathbf{p}_s^{\mathrm{tx}}
\right\|_2,
\label{eq:prior_distance}
\end{equation}
where \(\mathbf{p}_s^{\mathrm{tx}}\) denotes the AP position. The corresponding free-space path loss is obtained from the Friis equation:
\begin{equation}
L_{s,k}^{\mathrm{FS}}[i,j]
=
20\log_{10}\!\left(
\frac{
4\pi f_{c,s}
\max\!\left(d_{s,k}[i,j],d_{\min}\right)
}{c}
\right),
\label{eq:prior_fspl}
\end{equation}
where \(f_{c,s}\) is the carrier frequency, \(c\) is the speed of light, and \(d_{\min}>0\) prevents numerical instability near the AP.

Since free-space attenuation alone cannot represent directional transmission or obstruction effects, we augment the analytical prior with geometry-aware descriptors:
\begin{equation}
\begin{aligned}
\mathbf{Q}_{s,k}
=
\operatorname{Stack}\!\bigl(
&\widetilde{\log \mathbf{d}}_{s,k},
\widetilde{\mathbf{L}}_{s,k}^{\mathrm{FS}},
\mathbf{D}_{s,k}^{\mathrm{ori}},
\mathbf{B}_{s,k}^{\mathrm{LoS}},
\widetilde{\mathbf{N}}_{s,k}^{\mathrm{wall}},
\\[-1mm]
&\widetilde{\mathbf{D}}_{s,k}^{\mathrm{surf}},
\mathbf{B}_{s,k}^{\mathrm{in}}
\bigr),
\end{aligned}
\label{eq:encoded_propagation_prior}
\end{equation}
Here, \(\mathbf{d}_{s,k}\) and
\(\mathbf{L}_{s,k}^{\mathrm{FS}}\) collect
\(d_{s,k}[i,j]\) and \(L_{s,k}^{\mathrm{FS}}[i,j]\) over the receiver grid, respectively.
The orientation map \(\mathbf{D}_{s,k}^{\mathrm{ori}}\) measures the alignment between the AP pointing direction and the AP-to-receiver direction.
The binary mask \(\mathbf{B}_{s,k}^{\mathrm{LoS}}\) indicates whether the direct AP--receiver path is unobstructed.
\(\mathbf{N}_{s,k}^{\mathrm{wall}}\) counts wall intersections along that path,
\(\mathbf{D}_{s,k}^{\mathrm{surf}}\) measures the distance to the nearest scene surface, and
\(\mathbf{B}_{s,k}^{\mathrm{in}}\) identifies valid indoor receiver locations.

The height-specific descriptors are stacked and projected to the
RF-token embedding using a lightweight 3D convolutional encoder:
\begin{equation}
\boldsymbol{\Pi}_s
=
E_{\mathrm{prior}}
\left(
\operatorname{Stack}_{k=1}^{K}
\mathbf{Q}_{s,k}
\right)
\in
\mathbb{R}^{D\times K\times H_{\mathrm{lat}}\times W_{\mathrm{lat}}}.
\end{equation}
Here, \(E_{\mathrm{prior}}\) consists of three
\(3\times3\times3\) convolutional blocks followed by a
\(1\times1\times1\) projection. We denote its feature slice at
height \(z_k\) by
\(\boldsymbol{\Pi}_{s,k}
\in\mathbb{R}^{H_{\mathrm{lat}}\times W_{\mathrm{lat}}\times D}\).
\subsubsection{Multi-modal Conditioning}
In Stage~I, the RF prefix $\mathbf{c}_{s,k}^{\mathrm{pre}}$, the propagation prior $\boldsymbol{\Pi}_{s,k}$ and the text
embedding $\mathbf{e}_s$ are conditioned through separate pathways according to their
structures and semantics.
For receiver height \(z_k\), let
$\mathbf{X}_{s,k}^{(\ell)}
\in
\mathbb{R}^{T_{\mathrm{lat}}\times H_{\mathrm{lat}}
\times W_{\mathrm{lat}}\times D}$
denote the RF token embedding before the \(\ell\)-th DiT block.
The input RF tokens are then initialized as
\begin{equation}
\mathbf{X}_{s,k}^{(0)}
=
E_{\mathrm{in}}
\left(
\operatorname{Concat}_{\mathrm{ch}}
\left[
\mathbf{z}_{\sigma,s,k},
\mathbf{c}_{s,k}^{\mathrm{pre}}
\right]
\right)
+
\operatorname{Repeat}_{T_{\mathrm{lat}}}
\left(
\boldsymbol{\Pi}_{s,k}
\right),
\label{eq:stage1_input_tokens}
\end{equation}
where \(E_{\mathrm{in}}(\cdot)\) denotes the input projection and patchification
operator, \(\operatorname{Concat}_{\mathrm{ch}}(\cdot)\) denotes channel-wise
concatenation, and
\(\operatorname{Repeat}_{T_{\mathrm{lat}}}(\cdot)\) broadcasts the static prior
feature over the latent temporal axis.
The text embedding \(\mathbf{e}_s\) is then supplied to each DiT block through
cross-attention \cite{cross-ATTENTION}, providing scene-level semantic context throughout the
denoising process.

\subsubsection{Training Objective}

Given the noisy latent \(\mathbf{z}_{\sigma,s,k}\) and the three conditioning cues, the RF-adapted backbone predicts the complete RF latent:
\begin{equation}
\widehat{\mathbf{z}}_{0,s,k}
=
f_{\boldsymbol{\theta}_{\mathrm{RF}}}\!\left(
\mathbf{z}_{\sigma,s,k},
\sigma
\;\middle|\;
\mathbf{c}_{s,k}^{\mathrm{pre}},
\mathbf{e}_s,
\boldsymbol{\Pi}_{s,k}
\right),
\label{eq:rf_pretraining_prediction}
\end{equation}
where \(\boldsymbol{\theta}_{\mathrm{RF}}\) denotes the trainable parameters after RF-domain adaptation. Since prefix frames are already provided to the model, the denoising error is computed only on the unobserved portion of the trajectory:
\begin{equation}
\mathcal{L}_{\mathrm{pre}}
=
\mathbb{E}
\left[
w(\sigma)
\left\|
\left(
\mathbf{1}
-
\mathbf{B}_{s,k}^{\mathrm{pre}}
\right)
\odot
\left(
\widehat{\mathbf{z}}_{0,s,k}
-
\mathbf{z}_{0,s,k}
\right)
\right\|_{2,\mathrm{mean}}^{2}
\right],
\label{eq:rf_pretraining_objective}
\end{equation}
where \(w(\sigma)\) balances training samples across diffusion noise levels.

\subsection{Physical-to-RF Transfer}
\label{subsec:physical_to_rf}
As shown in \autoref{fig2}~(c), RFWM in Stage~II transfers the RF-aware backbone to physical-to-RF generation by conditioning it on \(\mathbf{C}_s^{\mathrm{obs}}\). Because these conditions differ in structure and physical role, RFWM integrates them through dedicated pathways and jointly predicts the RF trajectories at all \(K\) queried receiver heights.
Additionally, the prior used in the first stage does not fully constrain the fine-grained behavior of fields generated from physical observations. We therefore introduce six physics-guided regularizers that penalize local violations in attenuation, blockage, and spatial consistency. The following describes the structured conditioning pathways, physics-guided regularization, and joint multi-height prediction in detail.

\subsubsection{Structured Multimodal Conditioning}

The observations in \(\mathbf{C}_s^{\mathrm{obs}}\) differ in temporal
resolution, spatial structure and physical meaning. 
As illustrated in \autoref{fig2}~(c), RFWM preserves these distinctions by encoding each modality with a dedicated conditioning module.

\textbf{Dynamic physical control.}
The visual sequence identifies when and where the environment changes, while
the AP raster anchors those changes relative to the transmitter. We repeat the
static raster over time and concatenate it with the visual sequence
as $\mathbf{C}_s^{\mathrm{dyn}}
=
\operatorname{Concat}_{\mathrm{ch}}
\left(
\mathbf{V}_s,
\operatorname{Repeat}_{T}
\left(
\mathbf{A}_s
\right)
\right).$
A DiT-based ControlNet extracts frame-aligned features and injects projected
residuals into selected layers of the RF backbone, as expressed by
\begin{equation}
\mathbf{R}_{s,\mathrm{ctrl}}^{(\ell)}
=
P_{\mathrm{ctrl}}^{(\ell)}
\left(
F_{\mathrm{ctrl}}^{(\ell)}
\left(
\mathbf{C}_s^{\mathrm{dyn}}
\right)
\right),
\qquad
\ell\in\mathcal{I}_{\mathrm{ctrl}},
\label{eq:controlnet_injection}
\end{equation}
where $F_{\mathrm{ctrl}}^{(\ell)}(\cdot)$ and $P_{\mathrm{ctrl}}^{(\ell)}(\cdot)$ represent the ControlNet operator and the MLP projection layer.
Notably, the
frame-aligned residual is shared across height branches. 

\textbf{Wireless conditioning.}
Receiver height is included in
\(\mathbf{w}_{s,k}\) and further expanded with Fourier features, as expressed by
\begin{equation}
\gamma(z_k)
=
\left[
\sin
\left(
2^m\pi z_k
\right),
\cos
\left(
2^m\pi z_k
\right)
\right]_{m=0}^{L_\gamma-1}, 
\end{equation}
where $L_\gamma$ denotes the number of frequency bands.
A two-layer wireless encoder produces a
height-specific wireless embedding:
\begin{equation}
\begin{aligned}
\mathbf{c}_{s,k}^{\mathrm{wire}}
&=
\mathbf{W}_2
\operatorname{SiLU}
\left(
\mathbf{W}_1
\operatorname{LN}
\left[
\mathbf{w}_{s,k},
\gamma(z_k)
\right]
\right),
\\
\mathbf{h}_{\sigma,s,k}
&=
E_\sigma(\sigma)
+
\mathbf{c}_{s,k}^{\mathrm{wire}},
\end{aligned}
\label{eq:wireless_encoder}
\end{equation}
where \(\operatorname{LN}(\cdot)\) denotes layer normalization,
\(\operatorname{SiLU}(\cdot)\) is the activation function,
\(\mathbf{W}_1\) and \(\mathbf{W}_2\) are the weight matrices of the two
linear layers.

\textbf{Geometry and semantic conditioning.}
We voxelize the static geometry \(\mathbf{M}_s\) in the latent space and encode
it with a mesh encoder $E_{mesh}(\cdot)$, expressed by 
\begin{equation}
\begin{aligned}
\mathbf{F}_s^{\mathrm{mesh}}
&=
E_{\mathrm{mesh}}
\left(
\operatorname{Vox}
\left(
\mathbf{M}_s
\right)
\right),
\\
\mathbf{f}_{s,k}^{\mathrm{mesh}}
&=
\operatorname{Slice}_{k}
\left(
\mathbf{F}_s^{\mathrm{mesh}}
\right).
\end{aligned}
\label{eq:mesh_encoder}
\end{equation}
The slice \(\mathbf{f}_{s,k}^{\mathrm{mesh}}\) is repeated over latent time
and added to the tokens at height \(z_k\).  
The text embedding \(\mathbf{e}_s\) continues to provide
scene-level context through cross-attention, complementing the local geometric
features with global semantic information.

The injection of all physical conditions can
be summarized as
\begin{equation}
\begin{aligned}
\mathbf{X}_{s,k}^{(0)}
&=
E_{\mathrm{in}}
\left(
\mathbf{z}_{\sigma,s,k}
\right)
+
\operatorname{Repeat}_{T_{\mathrm{lat}}}
\left(
\mathbf{f}_{s,k}^{\mathrm{mesh}}
\right),
\\
\widetilde{\mathbf{X}}_{s,k}^{(\ell)}
&=
\mathbf{X}_{s,k}^{(\ell)}
+
\mathbb{I}
\left[
\ell\in\mathcal{I}_{\mathrm{ctrl}}
\right]
\mathbf{R}_{s,\mathrm{ctrl}}^{(\ell)},
\\
\mathbf{X}_{s,k}^{(\ell+1)}
&=
\mathcal{B}_{\mathrm{RF}}^{(\ell)}
\left(
\widetilde{\mathbf{X}}_{s,k}^{(\ell)};
\mathbf{h}_{\sigma,s,k},
\mathbf{e}_s
\right).
\end{aligned}
\label{eq:stage2_condition_injection}
\end{equation}
where $\mathcal{B}_{\mathrm{RF}}^{(\ell)}(\cdot)$ stands for the $\ell$-th DiT block.
The complete conditional denoising process is
\begin{equation}
\widehat{\mathbf{Z}}_{0,s}
=
f_{\boldsymbol{\theta}}
\left(
\mathbf{Z}_{\sigma,s},
\sigma
\;\middle|\;
\mathbf{C}_s^{\mathrm{dyn}},
\left\{
\mathbf{c}_{s,k}^{\mathrm{wire}}
\right\}_{k=1}^{K},
\mathbf{F}_s^{\mathrm{mesh}},
\mathbf{e}_s
\right),
\label{eq:stage2_prediction}
\end{equation}
where the RF-backbone parameters in \(\boldsymbol{\theta}\) are initialized
from \(\boldsymbol{\theta}_{\mathrm{RF}}\).

\subsubsection{Physics-Guided Regularization}
\label{subsubsec:physics_regularization}

\begin{figure}[t]
    \centering
    \includegraphics[width=\columnwidth]{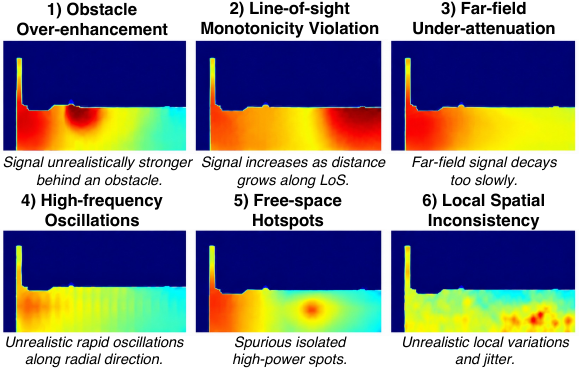}
    \caption{Representative physical inconsistencies in generated RF fields:
    obstacle over-enhancement, LoS monotonicity violation, far-field
    under-attenuation, high-frequency radial oscillations, free-space
    hotspots, and local spatial inconsistency.}
    \label{fig:bad_cases}
\vspace{-4mm}
\end{figure}
The Friis-guided prior used in Stage~I primarily captures coarse and large-scale propagation trends. However, it does not explicitly constrain fine-grained local propagation behavior, and the generated RF fields may still exhibit physically implausible artifacts. 
As illustrated in \autoref{fig:bad_cases}, we categorize these failure cases into six representative types.
To suppress these failure modes, we introduce six complementary physics-guided regularizers, each formulated as a soft penalty for the corresponding physical violation as follows.

Since the VAE-decoder predicts
\(\widehat{\mathbf{Y}}_{s,k}\), we first recover the corresponding RSS
trajectory through the inverse-rendering operator $\Phi^{-1}$ as 
$\widehat{\mathbf{R}}_{s,k}
=
\Phi^{-1}
\left(
\widehat{\mathbf{Y}}_{s,k}
\right)
\in
\mathbb{R}^{T\times H\times W}.$ For a receiver location
\(\mathbf{r}_{k,i,j}=(x_i,y_j,z_k)\), the predicted RSS is defined as
$\widehat{\rho}_{s,t,k}
\left(
\mathbf{r}_{k,i,j}
\right)
=
\widehat{\mathbf{R}}_{s,k}[t,i,j].$
To simplify notation, we omit the indices \((s,t,k)\) below and write
\(\widehat{\rho}(\mathbf{r})\), where a larger value indicates a stronger
received signal. We further define \([a]_{+}=\max(a,0)\), such that each
one-sided penalty is activated only when its corresponding physical relation
is violated.

 Let
\(\mathcal{P}_{\mathrm{obs}}\) contain ordered front--behind pairs
\((\mathbf{r}_{f},\mathbf{r}_{b})\) across the same obstacle. Let
\(\mathcal{P}_{\mathrm{LoS}}\) contain ordered near--far pairs
\((\mathbf{r}_{i},\mathbf{r}_{j})\) on the same unobstructed ray, where
$d_s(\mathbf{r}_{j})>d_s(\mathbf{r}_{i}),
\qquad
d_s(\mathbf{r})
=
\left\|
\mathbf{r}
-
\mathbf{p}^{\mathrm{tx}}_s
\right\|_2.$
Finally,
\(\mathcal{P}_{\mathrm{far}}\subseteq\mathcal{P}_{\mathrm{LoS}}\)
contains pairs in the far-field region.

\textbf{Obstacle attenuation.}
For each
\((\mathbf{r}_f,\mathbf{r}_b)\in\mathcal{P}_{\mathrm{obs}}\),
the RSS behind the obstacle should be lower than that in front by at least
\(m_{\mathrm{obs}}>0\), as expressed by
\begin{equation}
\mathcal{L}_{\mathrm{obs}}
=
\left\langle
\left[
\widehat{\rho}(\mathbf{r}_b)
-
\widehat{\rho}(\mathbf{r}_f)
+
m_{\mathrm{obs}}
\right]_+
\right\rangle_{\mathcal{P}_{\mathrm{obs}}}.
\label{eq:obstacle_attenuation_loss}
\end{equation}

\textbf{LoS monotonicity.}
For
\((\mathbf{r}_i,\mathbf{r}_j)\in\mathcal{P}_{\mathrm{LoS}}\) with
\(d_s(\mathbf{r}_j)>d_s(\mathbf{r}_i)\), we penalize an excessive positive
radial slope:
\begin{equation}
\mathcal{L}_{\mathrm{mon}}
=
\left\langle
\left[
\frac{
\widehat{\rho}(\mathbf{r}_j)
-
\widehat{\rho}(\mathbf{r}_i)
}{
d_s(\mathbf{r}_j)
-
d_s(\mathbf{r}_i)
}
-
m_{\mathrm{mon}}
\right]_+
\right\rangle_{\mathcal{P}_{\mathrm{LoS}}},
\label{eq:distance_monotonicity_loss}
\end{equation}
where \(m_{\mathrm{mon}}\geq0\) tolerates moderate multipath fluctuations.

\textbf{Far-field attenuation.}
Monotonicity alone does not ensure sufficient decay at large distances. We
therefore compare the predicted attenuation rate with the Friis reference:
\begin{equation}
\begin{aligned}
\widehat{\alpha}_{ij}
&=
\frac{
\widehat{\rho}(\mathbf{r}_i)
-
\widehat{\rho}(\mathbf{r}_j)
}{
\log_{10}
\!\left(
d_s(\mathbf{r}_j)/d_s(\mathbf{r}_i)
\right)
},
\\
\alpha_{ij}^{\mathrm{FS}}
&=
\frac{
\rho^{\mathrm{FS}}(\mathbf{r}_i)
-
\rho^{\mathrm{FS}}(\mathbf{r}_j)
}{
\log_{10}
\!\left(
d_s(\mathbf{r}_j)/d_s(\mathbf{r}_i)
\right)
},
\\
\mathcal{L}_{\mathrm{far}}
&=
\left\langle
\left[
\alpha_{ij}^{\mathrm{FS}}
-
\widehat{\alpha}_{ij}
-
m_{\mathrm{far}}
\right]_+^2
\right\rangle_{\mathcal{P}_{\mathrm{far}}},
\end{aligned}
\label{eq:far_field_loss}
\end{equation}
where
\(\rho^{\mathrm{FS}}(\mathbf{r})=-L^{\mathrm{FS}}(\mathbf{r})\)
is the Friis-based RSS trend, and \(m_{\mathrm{far}}\geq0\) allows moderate
deviation from this reference.

\textbf{Radial smoothness.}
Let \(\mathcal{Q}_{\mathrm{LoS}}\) contain equally spaced receiver triplets
\((\mathbf{r}_{i-1},\mathbf{r}_i,\mathbf{r}_{i+1})\) along unobstructed AP
rays. We suppress unsupported high-frequency oscillations using the
second-order radial difference:
\begin{equation}
\mathcal{L}_{\mathrm{rad}}
=
\left\langle
\left|
\widehat{\rho}(\mathbf{r}_{i+1})
-
2\widehat{\rho}(\mathbf{r}_i)
+
\widehat{\rho}(\mathbf{r}_{i-1})
\right|
\right\rangle_{\mathcal{Q}_{\mathrm{LoS}}}.
\label{eq:radial_smoothness_loss}
\end{equation}

\textbf{Free-space hotspot suppression.}
Let \(\mathcal{F}\) be the set of free-space receiver cells and
\(\mathcal{N}(\mathbf{r})\) the neighbors of \(\mathbf{r}\). With
\[
\overline{\rho}_{\mathcal{N}}(\mathbf{r})
=
\frac{1}{|\mathcal{N}(\mathbf{r})|}
\sum_{\mathbf{u}\in\mathcal{N}(\mathbf{r})}
\widehat{\rho}(\mathbf{u}),
\]
we penalize isolated high-power peaks:
\begin{equation}
\mathcal{L}_{\mathrm{hot}}
=
\left\langle
\left[
\widehat{\rho}(\mathbf{r})
-
\overline{\rho}_{\mathcal{N}}(\mathbf{r})
-
m_{\mathrm{hot}}
\right]_+
\right\rangle_{\mathcal{F}},
\label{eq:hotspot_loss}
\end{equation}
where \(m_{\mathrm{hot}}\geq0\) preserves legitimate local variation.

\textbf{Local spatial consistency.}
Let \(\mathcal{E}\) contain neighboring receiver-cell pairs
\((\mathbf{r},\mathbf{u})\). We encourage local coherence through
\begin{equation}
\mathcal{L}_{\mathrm{spa}}
=
\left\langle
\omega_{\mathbf{r},\mathbf{u}}
\left|
\widehat{\rho}(\mathbf{r})
-
\widehat{\rho}(\mathbf{u})
\right|
\right\rangle_{\mathcal{E}},
\label{eq:spatial_consistency_loss}
\end{equation}
where \(\omega_{\mathbf{r},\mathbf{u}}\in[0,1]\) is reduced across scene
surfaces or changes in LoS state, preventing smoothing across
geometry-induced propagation boundaries.
Notably, $m_{\mathrm{obs}},\;
m_{\mathrm{mon}},\;
m_{\mathrm{far}},\;
m_{\mathrm{hot}}$
are summarized from the failure cases after the training in Stage~I.

\subsubsection{Cross-Height RF Prediction and Training Objective}
RF trajectories at neighboring receiver heights exhibit correlated temporal and
vertical structure. RFWM thus models all \(K\) heights jointly rather than treating
them as independent prediction tasks. Let
\(\mathbf{X}_{s,k}^{(L)}\) denote the final DiT feature at height \(z_k\).
The features from all height branches are stacked and provided to each
height-specific output head, as expressed by
\begin{equation}
\mathbf{X}_s^{(L)}
=
\operatorname{Stack}_{k=1}^{K}
\left(
\mathbf{X}_{s,k}^{(L)}
\right),
\qquad
\widehat{\mathbf{z}}_{0,s,k}
=
H_k
\left(
\mathbf{X}_s^{(L)}
\right),
\label{eq:cross_height_heads}
\end{equation}
where \(H_k(\cdot)\) predicts the clean RF latent at  \(z_k\).
The predicted latents are decoded and stacked, yielding the complete
spatiotemporal RF field in one forward pass.

Stage~II is trained with a denoising loss over all heights and a vertical
consistency loss between adjacent heights:
\begin{equation}
\begin{aligned}
\mathcal{L}_{\mathrm{diff}}
&=
\mathbb{E}
\left[
\frac{w(\sigma)}{K}
\sum_{k=1}^{K}
\left\|
\widehat{\mathbf{z}}_{0,s,k}
-
\mathbf{z}_{0,s,k}
\right\|_{2,\mathrm{mean}}^2
\right],
\\
\mathcal{L}_{z}
&=
\mathbb{E}
\left[
\frac{1}{K-1}
\sum_{k=1}^{K-1}
\left\|
\Delta\widehat{\mathbf{z}}_{0,s,k}
-
\Delta\mathbf{z}_{0,s,k}
\right\|_1
\right],
\end{aligned}
\label{eq:stage2_prediction_losses}
\end{equation}
where
\(\Delta\widehat{\mathbf{z}}_{0,s,k}
=
\widehat{\mathbf{z}}_{0,s,k+1}
-
\widehat{\mathbf{z}}_{0,s,k}\)
and
\(\Delta\mathbf{z}_{0,s,k}
=
\mathbf{z}_{0,s,k+1}
-
\mathbf{z}_{0,s,k}\).
The complete training loss is
\begin{equation}
\mathcal{L}_{\mathrm{transfer}}
=
\mathcal{L}_{\mathrm{diff}}
+
\lambda_z\mathcal{L}_{z}
+
\mathcal{L}_{\mathrm{phy}},
\label{eq:transfer_loss}
\end{equation}
where \(\lambda_z\) controls the vertical consistency term and
\(\mathcal{L}_{\mathrm{phy}}\) denotes the weighted sum of the six physics-guided regularization terms.

\section{Experiments}
In this section, 
we evaluate RFWM on the constructed benchmark under both ID and OOD settings. We  compare RFWM with both current scene-specific and cross-scene methods in terms of reconstruction accuracy, generalization, and rendering efficiency. Finally, we assess the physical fidelity of the generated RF fields and conduct ablation studies to illustrate the effectiveness of the proposed physical regularizers.
\subsection{Experimental Setup}
\subsubsection{Datasets and Implementation}
We construct a physical-to-RF benchmark comprising 7,715 synchronized
physical--RF sequences following \cite{park2026diffusion}. Specifically, we first construct 115  3D environments from the large-scale 3D-FRONT dataset \cite{3d-front}. We then use the human
locomotion algorithm DIMOS \cite{DIMOS} to generate human walking trajectories within these
environments, introducing time-varying perturbations to the scene geometry.
Based on the resulting dynamic environments, AutoMS \cite{automs} is used to generate the
corresponding spatiotemporal RF fields. Each sample therefore consists of a
synchronized physical-observation sequence and its corresponding RF-field
sequence.
The training set contains 7,045 sequences from 101 scenes, which are further
divided into 23,139 training clips. The ID test set contains
293 sequences from 74 held-out routes in 37 scenes observed during training.
All sequences belonging to these routes are excluded from the training set.
The OOD test set contains 377 sequences from 14 scenes
that are entirely absent from training, ensuring no scene overlap between the
training and OOD sets.
RFWM is trained on eight NVIDIA H100 GPUs with FSDP and BF16
precision. We use AdamW with learning rates of $1\times10^{-5}$
for the pretrained DiT backbone and $4.96\times10^{-5}$ for the
newly introduced modules. Stage~I is trained for 600k iterations,
followed by a 10k-step conditional warm-up and 50k-step
physical-to-RF transfer in Stage~II.
For the physics-guided regularizers, 
we assign relative weights of
$1.0$, $0.5$, $1.0$, $0.2$, $0.5$, and $0.2$ to 
$\lambda_{\mathrm{obs}},
 \lambda_{\mathrm{mon}},
 \lambda_{\mathrm{far}},
\lambda_{\mathrm{rad}},
\lambda_{\mathrm{hot}}$
and
$
\lambda_{\mathrm{spa}}$
respectively.

\begin{table*}[t]
\vspace{-4mm}
\centering
\caption{RF-field modeling performance and average scene-level rendering time
under ID and OOD evaluation. 
Rendering time measures the average time required to generate one complete 3D RF-field frame over all queried spatial locations.}
\label{tab:main_results}

\setlength{\tabcolsep}{3pt}

\begin{tabular*}{\textwidth}{
@{\extracolsep{\fill}}
lcccccccccc
@{}
}
\toprule
\multirow[c]{2}{*}[-1.2ex]{\makecell[c]{\textbf{Method}}}
&
\multirow[c]{2}{*}[-1.2ex]{\makecell[c]{\textbf{Modeling}\\\textbf{Paradigm}}}
&
\multirow[c]{2}{*}[-1.2ex]{\makecell[c]{\textbf{Rendering}\\\textbf{Time (s)}}}
& \multicolumn{4}{c}{\textbf{ID}}
& \multicolumn{4}{c}{\textbf{OOD}} \\
\cmidrule(lr){4-7}
\cmidrule(lr){8-11}
&
&
&
\makecell[c]{MSE\\(dB) $\downarrow$}
& \makecell[c]{PSNR\\(dB) $\uparrow$}
& SSIM $\uparrow$
& LPIPS $\downarrow$
& \makecell[c]{MSE\\(dB) $\downarrow$}
& \makecell[c]{PSNR\\(dB) $\uparrow$}
& SSIM $\uparrow$
& LPIPS $\downarrow$ \\
\midrule

NeRF$^2$
& \multirow{4}{*}{\makecell[c]{One model \\per scene}}
& 4.3619
& -8.6140
& 11.2240
& 0.5296
& 0.5032
& -6.7045
& \underline{6.9664}
& 0.3859
& 0.6974 \\

WRF-GS
&
& 3.0148
& -11.5650
& \underline{14.6860}
& \underline{0.6895}
& \underline{0.3016}
& -6.3990
& 6.5427
& \underline{0.4046}
& \underline{0.6774} \\

RadCloudSplat
&
& 0.2910
& -11.2720
& 13.7130
& 0.5194
& 0.3863
& \underline{-6.7582}
& 6.8754
& 0.3245
& 0.6825 \\

RFCanvas
&
& 0.7216
& \underline{-12.8535}
& 13.5549
& 0.6661
& 0.3897
& -3.1898
& 3.2129
& 0.1882
& 0.9072 \\

\midrule

\textbf{RFWM}
& \makecell[c]{One model across\\dynamic scenes}
& 0.7697
& \textbf{-18.4664}
& \textbf{22.3067}
& \textbf{0.8145}
& \textbf{0.0698}
& \textbf{-10.6225}
& \textbf{11.1471}
& \textbf{0.6427}
& \textbf{0.3406} \\

\bottomrule
\vspace{-4mm}
\end{tabular*}
\end{table*}
\begin{figure*}[t]
    \vspace{-2mm}
    \centering
    \includegraphics[width=\textwidth]{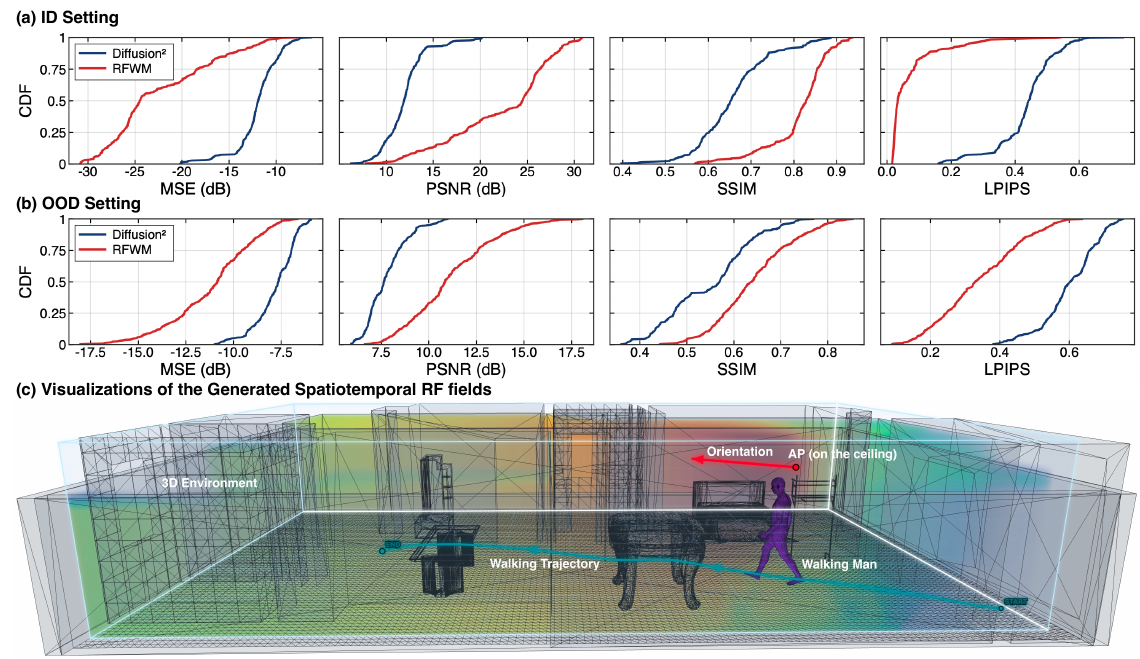}
    \caption{Illustration of performance comparison with Diffusion$^2$. (a)--(b) Empirical CDFs of MSE,
    PSNR, SSIM, and LPIPS on the ID and OOD test sets, respectively.
    (c) Example of an RFWM-generated spatiotemporal RF field aligned with the
    3D environment, AP configuration, and human trajectory.}
    \label{fig3}
    \vspace{-4mm}
\end{figure*}
\subsubsection{Baselines}
We compare RFWM with five representative learning-based RF-field modeling methods. Specifically, NeRF$^2$~\cite{zhao2023nerf2}, WRF-GS~\cite{wen2025wrfgs}, RadCloudSplat~\cite{wang2026radcloudsplat}, and RFCanvas~\cite{chen2024rfcanvas} follow a scene-specific fitting paradigm, requiring a separate model to be optimized for each environment. Diffusion$^2$~\cite{park2026diffusion} instead uses a shared model, but its original formulation generates a 2D RF heatmap at a fixed receiver height. For a fair 3D comparison, we retrain Diffusion$^2$ on our benchmark with the queried receiver height as an additional condition and stack the generated height-specific slices to construct the complete 3D RF field.

\subsubsection{Metrics}
We evaluate RF-field generation using four complementary metrics: MSE, PSNR, SSIM, and LPIPS.
MSE and PSNR are computed in the RSS domain to directly quantify the fidelity of the predicted RF values.
Let \(S\) denote the number of test samples, and let
\(\widehat{\mathbf{R}}_s\) and \(\mathbf{R}_s\) denote the generated and
ground-truth RSS trajectories of sample \(s\), respectively, with RSS values
normalized to \([0,1]\).
With \(N\) denoting the number of sampled RF points in each trajectory,
MSE and PSNR are computed as
\begin{equation}
\begin{aligned}
\operatorname{MSE}_{\mathrm{dB}}
&=
10\log_{10}
\left(
\frac{1}{SN}
\sum_{s=1}^{S}
\left\|
\widehat{\mathbf{R}}_s-\mathbf{R}_s
\right\|_2^2
\right),
\\
\operatorname{PSNR}_{\mathrm{dB}}
&=
-\frac{10}{S}
\sum_{s=1}^{S}
\log_{10}
\left(
\frac{1}{N}
\left\|
\widehat{\mathbf{R}}_s-\mathbf{R}_s
\right\|_2^2
\right).
\end{aligned}
\label{eq:evaluation_metrics}
\end{equation}

SSIM and LPIPS are evaluated in the RGB domain to assess the structural
and perceptual similarity of the rendered RF fields.
Specifically, the generated and ground-truth RF fields are converted into
RGB representations using the same rendering operator.
SSIM measures structural similarity following~\cite{wang2004image}, while
LPIPS quantifies perceptual discrepancy using AlexNet
features~\cite{zhang2018unreasonable}.
Lower MSE and LPIPS, and higher PSNR and SSIM, indicate better performance.
\subsection{Results}

\subsubsection{Comparison with Scene-Specific Methods}

\autoref{tab:main_results} compares RFWM with four scene-specific RF-field
modeling methods under the ID and OOD settings. 
RFWM consistently achieves the best performance among these methods. On the ID set, it obtains
\(-18.4664\) dB MSE, \(22.3067\) dB PSNR, \(0.0698\) LPIPS, and \(0.8145\)
SSIM. Relative to the strongest scene-specific result for each metric, RFWM
reduces MSE by \(5.61\) dB and LPIPS by \(76.9\%\), while improving PSNR by
\(7.62\) dB and SSIM by \(0.1250\). The advantage remains substantial under
OOD evaluation, with improvements of \(3.86\) dB in MSE, \(4.18\) dB in PSNR,
\(49.7\%\) in LPIPS, and \(0.2381\) in SSIM over the best corresponding
baseline. These consistent gains show that RFWM learns a transferable
physical-to-RF mapping and is capable of generalizing to unseen scenes.

The reported rendering time in \autoref{tab:main_results} is the average time required to generate one
complete 3D RF-field frame over all spatial locations and queried receiver
heights.
The scene-specific baselines predict one
RF point per query and therefore require repeated inference to assemble the
full RF field. By contrast, RFWM generates the entire spatiotemporal RF field in a single inference. 
Despite its large model size of approximately \(8\)B parameters, RFWM completes scene-level rendering in only \(0.7697\)s, achieving a favorable balance between reconstruction accuracy and inference efficiency.

\begin{table}[t]
\vspace{-4mm}
\centering
\caption{Comparison with Diffusion$^2$ and ablation of physics-guided regularization on the complete ID and OOD test sets.}
\label{tab:physics_checkpoint_ablation}

\small
\setlength{\tabcolsep}{2.0pt}
\renewcommand{\arraystretch}{1.12}

\begin{tabular*}{\columnwidth}{
@{\extracolsep{\fill}}
clccccc
@{}
}
\toprule
\textbf{Setting}
& \textbf{Method}
& \makecell[c]{\textbf{Rendering}\\\textbf{Time (s)} $\downarrow$}
& \makecell[c]{MSE\\(dB) $\downarrow$}
& \makecell[c]{PSNR\\(dB) $\uparrow$}
& SSIM $\uparrow$
& LPIPS $\downarrow$ \\
\midrule

\multirow{3}{*}{\textbf{ID}}
& Diffusion$^2$
& 0.5092
& -11.44
& 11.95
& 0.66
& 0.44 \\

& \makecell[l]{RFWM\\w/o PR}
& \multirow[c]{2}{*}{0.7697}
& \underline{-17.81}
& \underline{21.45}
& \underline{0.79}
& \underline{0.08} \\

& \textbf{RFWM}
&
& \textbf{-18.47}
& \textbf{22.31}
& \textbf{0.81}
& \textbf{0.07} \\

\midrule

\multirow{3}{*}{\textbf{OOD}}
& Diffusion$^2$
& 0.5092
& -7.67
& 7.80
& 0.55
& 0.60 \\

& \makecell[l]{RFWM\\w/o PR}
& \multirow[c]{2}{*}{0.7697}
& \underline{-9.87}
& \underline{10.12}
& \underline{0.61}
& \underline{0.37} \\

& \textbf{RFWM}
&
& \textbf{-10.62}
& \textbf{11.15}
& \textbf{0.64}
& \textbf{0.34} \\

\bottomrule
\vspace{-6mm}
\end{tabular*}
\end{table}
\begin{figure}[!htb]
    \centering
    \vspace{-4mm}
    \includegraphics[width=\columnwidth]{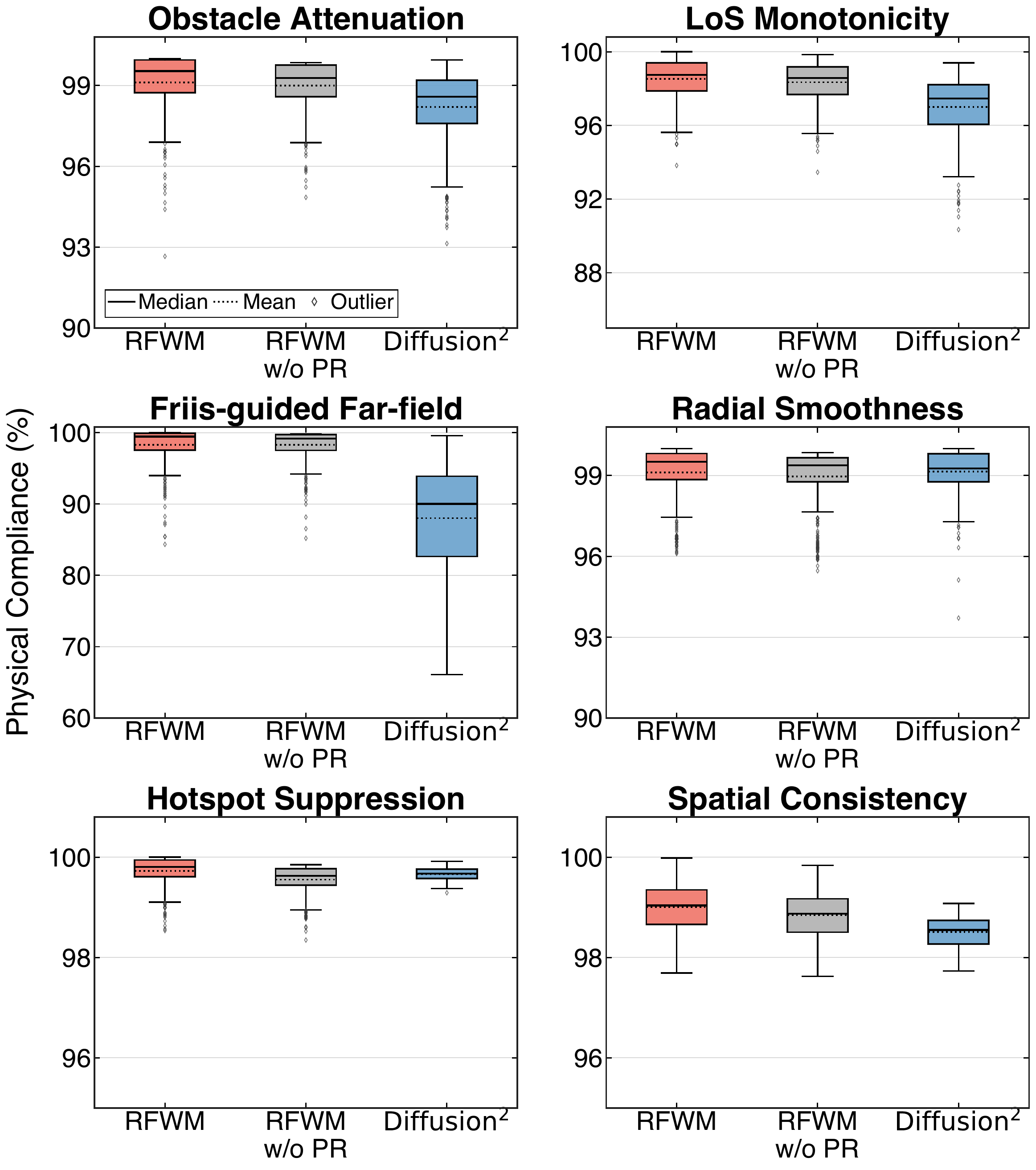}
     \vspace{-4mm}
    \caption{Physical compliance under the six propagation criteria used by the physics-guided regularizers. Box plots compare RFWM, RFWM without physics-guided regularization (w/o PR), and Diffusion$^2$.}
    \label{fig4}
    \vspace{-4mm}
\end{figure}
\subsubsection{Comparison with Scene-adaption Baseline}

As shown in \autoref{fig3}~(a)--(b), RFWM achieves consistently better performance compared with Diffusion$^2$ under both ID and OOD evaluation, yielding lower MSE and LPIPS together with higher PSNR and SSIM across most test samples. Its pronounced advantage on unseen scenes further validates the robustness and transferability of the learned physical-to-RF mapping of RFWM.
Moreover, \autoref{fig3}~(c) shows that RFWM generates a coherent spatiotemporal RF field aligned with the 3D environment, AP configuration, and human trajectory. It highlights the benefit of jointly modeling temporal evolution and cross-height dependencies instead of assembling independently generated RF slices.

\subsection{Physical Compliance and Ablation Study}
To assess physical compliance, we sample propagation relations from the
predictions of the models on the OOD set and calculate the
fraction that satisfy the corresponding physical constraint.
RFWM w/o PR means the model is trained without the six physics-guided regularizers.

As shown in \autoref{fig4}, RFWM achieves consistently stronger physical compliance across the six physical criteria. 
The most pronounced gain appears in far-field attenuation, where RFWM yields a higher and more concentrated compliance distribution than Diffusion$^2$, attributed to the Friis guidance used in both stages.
Removing PR degrades compliance across nearly all criteria, confirming that the six fine-grained constraints complement the coarse propagation guidance learned in Stage~I and effectively suppress fine-grained physical artifacts.
Additionally, the ablation results in \autoref{tab:physics_checkpoint_ablation} further show that this improved physical fidelity translates into better reconstruction quality. Compared with RFWM w/o PR, the full version improves MSE by \(0.66/0.75\) dB and PSNR by \(0.86/1.03\) dB on the ID/OOD sets, respectively.

\section{Conclusion}
This paper reformulated dynamic RF-field modeling as a \emph{physical-to-RF world transfer} problem and introduced \textbf{RFWM}, a physics-guided RF world model that generates spatiotemporal RF fields from physical-world observations without target-scene RF measurements. Through two-stage training, RFWM first learns RF dynamics from historical trajectories and then maps multimodal physical observations to RF fields, guided by propagation priors and physics-based regularization. Its cross-height heads efficiently generate complete RF fields at all queried receiver heights in a single forward pass. Experiments on our benchmark of 7,715 dynamic RF sequences from 115 environments show that RFWM reduces MSE over the strongest baseline by approximately \(7\)~dB on in-distribution scenes and \(3\)~dB on out-of-distribution scenes. By enabling accurate 3D dynamic RF-field generation in unseen environments without additional RF measurements, RFWM establishes a new paradigm for scalable and transferable RF modeling.

\newpage

\bibliographystyle{IEEEtran}
\bibliography{IEEEabrv,Reference}

\end{document}